\documentclass[twocolumn,showpacs,aps,prl,superscriptaddress,floatfix]{revtex4}
\usepackage{epsfig}

\input{babarsym}
\newcommand{\pmstat}[1]{\pm#1_{\mathrm{stat}}}
\newcommand{\pmsyst}[1]{\pm#1_{\mathrm{syst}}}
\newcommand{\pmff}[2]{{}^{+#1}_{-#2}{}_{\mathrm{FF}}}
\newcommand{\BtoXulnu}{\ensuremath{B\to X_u\ell\nu}}
\newcommand{\BtoXclnu}{\ensuremath{B\to X_c\ell\nu}}
\newcommand{\BtoXlnu}{\ensuremath{B\to X\ell\nu}}
\newcommand{\Btag}{\ensuremath{B_{\mathrm{tag}}}}
\newcommand{\Bztopimlnu}{\ensuremath{\Bz\to\pim\ellp\nu}}

\newcommand{\DorDstar}{\ensuremath{D^{(*)}}}
\newcommand{\BtoDorDstarlnu}{\ensuremath{B\to\DorDstar\ell\nu}}
\newcommand{\pell}{\ensuremath{\mathbf{p}_{\ell}}}
\newcommand{\BY}{\ensuremath{\theta_{BY}}}
\newcommand{\cosBY}{\ensuremath{\cos\BY}}
\newcommand{\Bpil}{\ensuremath{\theta_{B\pi\ell}}}
\newcommand{\cosBpil}{\ensuremath{\cos\Bpil}}
\newcommand{\EB}{\ensuremath{E_B}}
\newcommand{\EY}{\ensuremath{E_Y}}
\newcommand{\Epil}{\ensuremath{E_{\pi\ell}}}
\newcommand{\mB}{\ensuremath{m_B}}
\newcommand{\mY}{\ensuremath{m_Y}}
\newcommand{\mpil}{\ensuremath{m_{\pi\ell}}}
\newcommand{\mmB}{\ensuremath{m^2_B}}
\newcommand{\mmY}{\ensuremath{m^2_Y}}
\newcommand{\mmpil}{\ensuremath{m^2_{\pi\ell}}}
\newcommand{\pB}{\ensuremath{\mathbf{p}_B}}
\newcommand{\pY}{\ensuremath{\mathbf{p}_Y}}
\newcommand{\ppil}{\ensuremath{\mathbf{p}_{\pi\ell}}}
\newcommand{\ppi}{\ensuremath{\mathbf{p}_{\pi}}}
\newcommand{\Eres}{\ensuremath{E_{\mathrm{res}}}}
\newcommand{\phiB}{\ensuremath{\phi_B}}
\newcommand{\RR}{\ensuremath{\cos^2\phiB}}

\newcommand{\Nsig}{\ensuremath{N_{\mathrm{sig}}}}
\newcommand{\Nbkg}{\ensuremath{N_{\mathrm{bkg}}}}
\newcommand{\Ncmb}{\ensuremath{N_{\mathrm{cmb}}}}
\newcommand{\PDF}{\ensuremath{\mathcal{P}}}
\newcommand{\Psig}{\ensuremath{\PDF_{\mathrm{sig}}}}
\newcommand{\Pbkg}{\ensuremath{\PDF_{\mathrm{bkg}}}}
\newcommand{\Pcmb}{\ensuremath{\PDF_{\mathrm{cmb}}}}
\newcommand{\err}[2]{{}^{+#1}_{-#2}}
\newcommand{\tauBp}{\ensuremath{\tau_{\Bp}}}
\newcommand{\tauBz}{\ensuremath{\tau_{\Bz}}}
\newcommand{\DBF}{\ensuremath{\Delta\BR}}
\newcommand{\mES}{\ensuremath{m_{\mathrm{ES}}}}
\newcommand{\rhom}{\ensuremath{\rho^-}}

\newcommand{\pmiss}{\ensuremath{p_{\mathrm{miss}}}}
\newcommand{\mmiss}{\ensuremath{m_{\mathrm{miss}}^2}}
\newcommand{\etal}{\textit{et al.}}

\newcommand{\BABARPubYear}{06}
\newcommand{\BABARPubNumber}{040}
\newcommand{\SLACPubNumber}{11966}
\newcommand{\LANLNumber}{0607089}

\begin{document}

\preprint{\babar-PUB-\BABARPubYear/\BABARPubNumber} 
\preprint{SLAC-PUB-\SLACPubNumber} 

\begin{flushleft}
\babar-PUB-\BABARPubYear/\BABARPubNumber\\
SLAC-PUB-\SLACPubNumber\\
hep-ex/\LANLNumber\\
\end{flushleft}

\title{\large\bf\boldmath
  Measurement of the $\Btopilnu$ Branching Fraction  and Determination of $\Vub$\\
  with Tagged $B$ Mesons}

%
\author{B.~Aubert}
\author{R.~Barate}
\author{M.~Bona}
\author{D.~Boutigny}
\author{F.~Couderc}
\author{Y.~Karyotakis}
\author{J.~P.~Lees}
\author{V.~Poireau}
\author{V.~Tisserand}
\author{A.~Zghiche}
\affiliation{Laboratoire de Physique des Particules, F-74941 Annecy-le-Vieux, France }
% ...
\collaboration{The \babar\ Collaboration}
\noaffiliation

\date{\today}

\begin{abstract}
  We report a measurement of the $\Btopilnu$ branching fraction based on
  211\invfb\ of data collected with the \babar\ detector.
  We use samples of $\Bz$ and $\Bp$ mesons tagged by
  a second $B$ meson reconstructed in a semileptonic or hadronic decay,
  and combine the results assuming isospin symmetry to obtain
  $\BR(\Bztopimlnu) = (1.33\pmstat{0.17}\pmsyst{0.11})\times10^{-4}$.
  We determine the magnitude of the Cabibbo-Kobayashi-Maskawa matrix
  element \Vub\ by combining the partial branching fractions measured
  in ranges of the momentum transfer squared
  and theoretical calculations of the form factor.
  Using a recent lattice QCD calculation, we find
  $\Vub=(4.5\pmstat{0.5}\pmsyst{0.3}\pmff{0.7}{0.5})\times10^{-3}$,
  where the last error is due to the normalization of the form factor.
\end{abstract}

\pacs{13.20.He,                 
      12.15.Hh,                 
      12.38.Qk,                 
      14.40.Nd}                 

\maketitle  


The magnitude of the Cabibbo-Kobayashi-Maskawa matrix~\cite{ckm}
	element $V_{ub}$ is a critical constraint on the Unitarity Triangle.
Our knowledge of \Vub\ comes from measurements of the $b\to u\ell\nu$
	decay rate, where the hadronic system in the final state can be  
	reconstructed either inclusively or exclusively.
The precisions are limited by the uncertainties in the 
	non-perturbative QCD calculations
	that are used to extract \Vub\ from the measured decay rates.
It is therefore crucial to pursue both the inclusive and exclusive approaches,
	which rely on different theoretical methods,
	and to test their consistency.


The rate of the exclusive decay $\Btopilnu$
	($\ell = e$ or $\mu$)
	is related to $\Vub$ through the form factor
	$f_+(q^2)$, where $q^2$ is the momentum transfer squared.
Measurements of the $\Btopilnu$ branching fraction have been
	reported by CLEO~\cite{cleo_vub}, \babar~\cite{babar_vub},
	and Belle~\cite{belle_vub}.
In this Letter, we report a measurement in which $\Btopilnu$ decays
	are searched for in $\Upsilon(4S)\to\BB$ events that are
	identified by reconstruction of the second $B$ meson (\Btag).
The technique, which was also used in~\cite{belle_vub},
	allows us to constrain the kinematics,
	reduce the combinatorics, and determine the charge of the signal $B$.
The result is an improved signal purity at the expense of the
	efficiency compared with the traditional measurements in which only
	the signal $B$ meson is reconstructed.
We perform two analyses in which $\Btag$ is reconstructed
	in semileptonic and hadronic decays, respectively,
	and combine the measured partial branching fractions $\Delta\BR$
	in ranges of $q^2$
  	with the recent form-factor
	calculations~\cite{lcsr:pi,lqcd:hpqcd,lqcd:fnal,lqcd:ape}
	to determine \Vub.


The measurement uses a sample of approximately 232 million \BB\ pairs,
  corresponding to an integrated luminosity of {211\invfb},
  recorded near the $\Upsilon(4S)$ resonance
  with the \babar\ detector~\cite{babar}
  at the PEP-II asymmetric-energy $\epem$ storage rings.
We use a detailed Monte Carlo (MC) simulation to estimate
  the signal efficiency and the signal and background distributions.


In the first analysis, we reconstruct $\Btag$
  in the semileptonic decay $\BtoDorDstarlnu$.
We reconstruct $\Dz$ mesons in $\Km\pip$, $\Km\pip\pip\pim$,
  $\Km\pip\piz$, and $\KS\pip\pim$ decays, and $\Dp$ mesons
  in $\Km\pip\pip$ decays~\cite{charge}.
The $D$ mass resolution ($\sigma$) is between $4.6$ and $12.9\mev$
  depending on the decay channel.
The mass of the $D$ candidate is required to be within $2.6\sigma$
  and $3.0\sigma$ of the expected value for the $\Bz$ and $\Bp$
  channels, respectively.
We also use a sideband sample, in which the $D$ candidate mass is 
  more than $3\sigma$ away from the nominal value, for subtracting the 
  combinatoric background.
We reconstruct $\Dstarp$ mesons in $\Dz\pip$ and $\Dp\piz$ decays.
The mass difference between the $\Dstar$ and $D$ is required
  to be within $3\mev$ of the expected value~\cite{pdg2004}.
The reconstructed $D$ and $\Dstar$ candidates are paired with
  a charged lepton with a center-of-mass (c.m.)\ momentum
  $|\pell|>0.8\gev$
  to form a $Y = \DorDstar\ell$ system.
If the $D$ decay contains a charged kaon, the lepton must have the
  same charge as the kaon.
The lepton and the $D$ meson are required to originate from a common
  vertex.
Assuming that only a massless neutrino escaped detection,
  we calculate the cosine of the
  angle between the $B$ and $Y$ momenta as
$
\cosBY = (2\EB\EY - \mmB - \mmY)/(2|\pB||\pY|),
$
where $\mB$, $\mY$, $\EB$, $\EY$, $\pB$, $\pY$ refer
to the masses, c.m.\ energies, and momenta of the $B$ and $Y$, respectively.
For background events, $\cosBY$ does not correspond to the cosine of a 
	physical angle and can extend outside $\pm1$.
We apply a loose selection of $|\cosBY|<5$ at this stage.


After identifying the $\Btag$ meson, we require the remaining particles
	in the event to be consistent with a $\Btopilnu$ decay.
Charged tracks that are not identified as a lepton or a kaon are considered
	charged pion candidates.
Neutral pion candidates are formed from pairs of photon candidates with
	invariant mass between 115 and 150\mev.
For the $\Bz$ channel,
  the lepton must have $|\pell|>0.8\gev$,
  and its charge must be opposite to that of the charged pion.
The lepton charge must be opposite to that of the $\Btag$
  for the $\Bp$ channel.
We reject the lepton candidate if, when combined with an oppositely-charged
  track, it is consistent with a $\jpsi\to\ellp\ellm$ decay or a 
  photon conversion.
Once the signal $B$ candidate is identified,
  we require that the event contain no other charged particles
  and small total c.m.\ energy $\Eres$ of the residual neutral particles.
In measuring $\Eres$, we remove the neutral candidates that are
  consistent with coming from a $\Dstar\to D\piz$ or $D\gamma$
  decay, bremsstrahlung from an electron, or beam-related background.
We require $\Eres<70\mev$ for the $\Bz$ channel and
  $\Eres<250\mev$ for the $\Bp$ channel, the latter being relaxed to
  allow for additional photons from decays of $\Dstarz$ and higher
  resonances.
We calculate the cosine of the
angle between the $B$ and $\pi\ell$ momenta as
$
\cosBpil = (2\EB\Epil - \mmB - \mmpil)/(2|\pB||\ppil|),
$
where $\mpil$, $\Epil$, $\ppil$ are
the mass, c.m.\ energy, and momentum of the $\pi\ell$ system, respectively.
We require $|\cosBpil|<5$.


Ignoring the small c.m.\ momentum of the $B$ meson,
  the invariant mass squared of the lepton-neutrino system
  in a $\Btopilnu$ decay can be inferred as
  $ q^2 = (\mB - E_{\pi})^2 - |\ppi|^2 $,
  where $E_{\pi}$ and $\ppi$ are the c.m.\ energy and momentum
  of the pion.
We divide the data into three bins:
  $q^2<8\gev^2$, $8<q^2<16\gev^2$, and $q^2>16\gev^2$.
We use simulated $\Btopilnu$ events to estimate and to correct for
  the small ($<8\%$) migration between the $q^2$ bins due to resolution,
  which is approximately $0.8\gev^2$ at $q^2=8\gev^2$ and improves with
  increasing $q^2$.

\begin{figure}
\epsfig{file=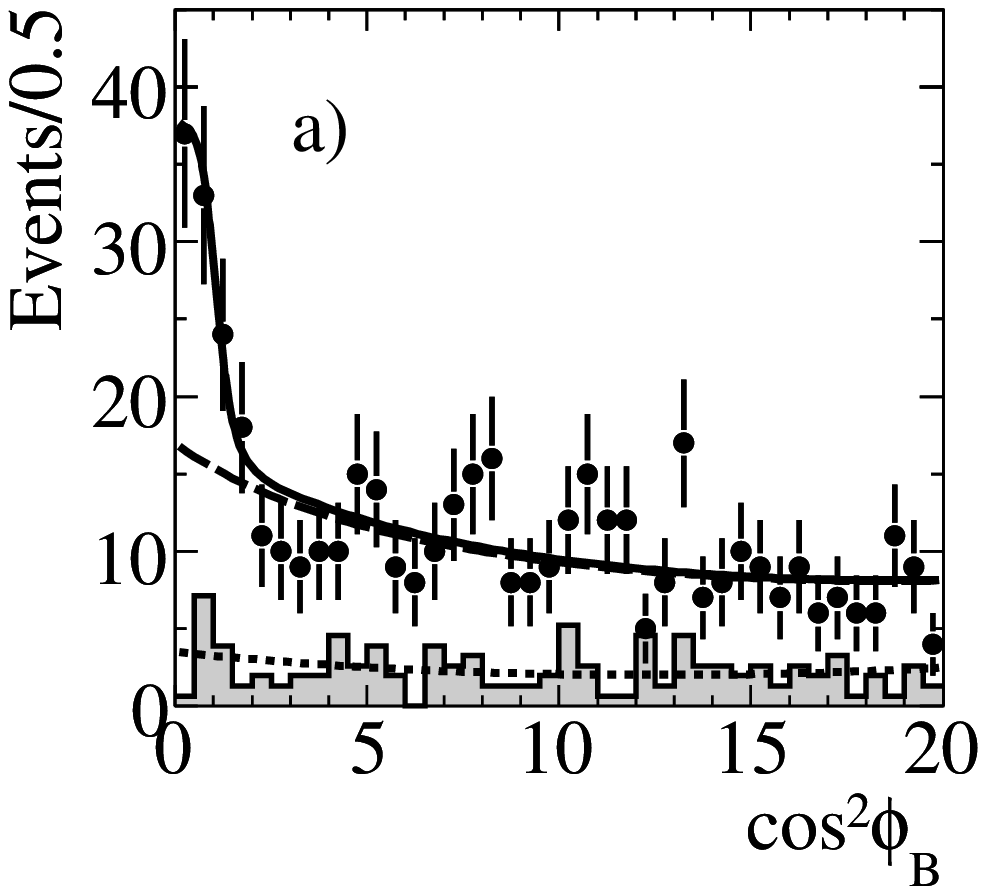,width=0.5\columnwidth}%
\epsfig{file=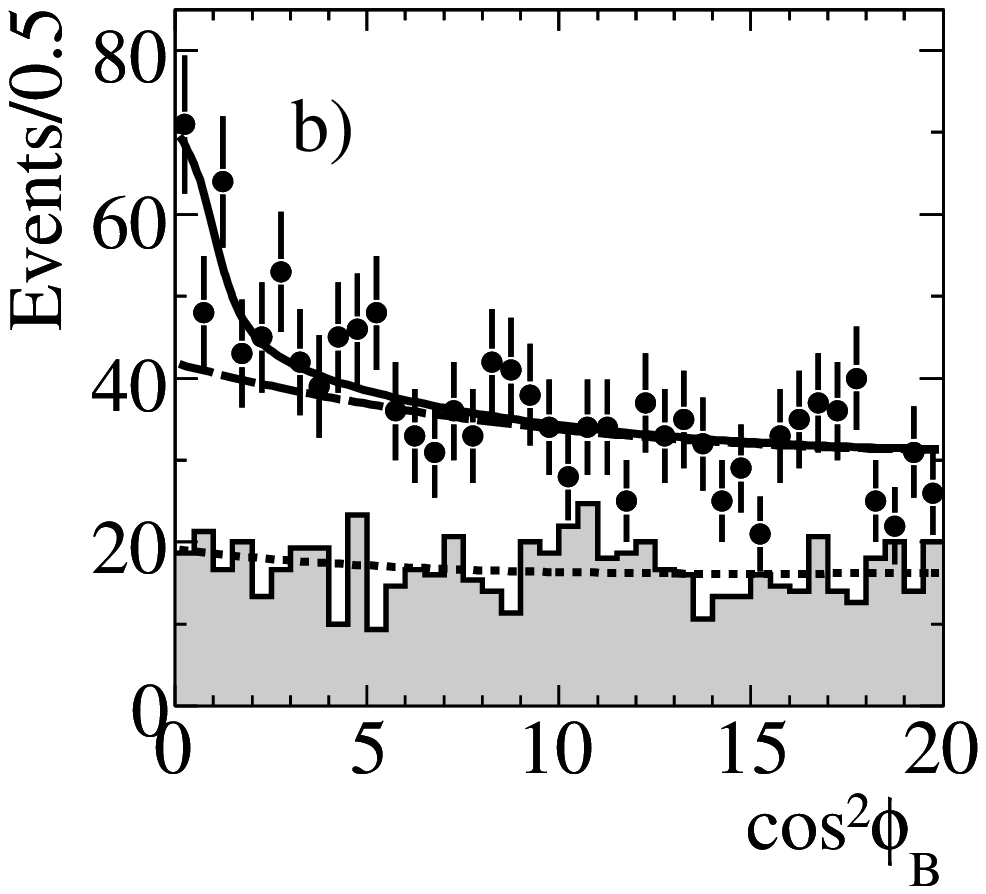,width=0.5\columnwidth}
\caption{\label{fig:slfit}
	Distributions of \RR\ of the a) $\Bz\to\pim\ellp\nu$
  and b) $\Bp\to\piz\ellp\nu$ candidates.
  The points with error bars and the shaded histograms are
  the data in the $D$ mass peak and sideband, respectively.
  The curves 
  are the fit results representing the total (solid),
  background (dashed), and `cmb' (dotted) components defined in the text.
  The fits were performed in bins of $q^2$, but the results shown
  are for the complete $q^2$ range.}
\end{figure}

Having identified the two $B$ mesons that decayed semileptonically,
  conservation of the total momentum determines
  the angle $\phiB$ between the direction of the 
  $B$ momenta and the plane defined by the $Y$ and $\pi\ell$ momenta:
  \begin{equation}
    \RR = \frac{\cos^2\BY + \cos^2\Bpil + 2\cosBY\cosBpil\cos\gamma}
	       {\sin^2\gamma},
  \end{equation}
  where $\gamma$ is the angle between the $Y$ and $\pi\ell$ momenta.
The variable $\RR$ satisfies $\RR\le1$ for correctly reconstructed
  signal events, and is broadly distributed for the background
  (see Fig.~\ref{fig:slfit}).
We use the $\RR$ distributions to extract the signal yield
  in the data in each $q^2$ bin.
We did not require stringent cuts on $\cosBY$ and $\cosBpil$
  because they are incorporated in $\RR$.
  

We express the data distribution as a sum of three
  contributions:
  $ dN/d\RR = \Nsig\Psig + \Nbkg\Pbkg + \Ncmb\Pcmb $,
  where $N_c$ and $\PDF_c$ are the number of events and the
  probability density function (PDF) for each category $c$,
  defined as the signal (sig), 
  background with correctly-reconstructed $D$ mesons (bkg),
  and other backgrounds (cmb).
The events in the $D$ mass sideband are also used in the fit
  to constrain the $\Ncmb\Pcmb$ term.
The PDF shapes are determined from the MC simulation.
The signal PDF is a combination of a smeared
  step function and an exponential tail.
The background PDFs are either an exponential
  plus constant or a second order polynomial.
The two data samples ($D$ mass peak and sideband)
  and the MC samples are used in an unbinned
  maximum likelihood fit that determines $\Nsig$, $\Nbkg$,
  $\Ncmb$, and the PDF parameters simultaneously.
Figure~\ref{fig:slfit} shows the fit results summed over the $q^2$ bins.
We find the signal yields and their statistical
  errors to be $57\err{13}{12}$ events and
  $92\err{26}{24}$ events for the $\Bz$ and $\Bp$
  channels, respectively.


We use simulated \Btopilnu\ events to estimate the signal
  efficiencies.
Control samples are used to derive corrections for the
  data-MC differences in the $\Btag$ reconstruction,
  charged and neutral particle reconstruction, and
  lepton identification.
The largest uncertainty comes from the $\Btag$ reconstruction
  efficiency, which is determined from a sample of 
  events in which two non-overlapping $\Btag$ candidates are
  reconstructed.
The efficiency correction factors for the $\Btag$ reconstruction
  are found to be $1.00\pm0.07$ and $0.99\pm0.02$ for the
  $\Bz$ and $\Bp$ channels, respectively.
The average signal efficiencies after the correction are
  $1.1\times10^{-3}$ for the $\Bz$ channel
  and $3.0\times10^{-3}$ for the $\Bp$ channel.
The latter is larger mainly because of the higher efficiency
  of reconstructing a $\Dz$ meson compared with a $\Dp$ or $\Dstarp$
  meson.


The measured branching fractions are summarized in Table~\ref{tab:bf}.
\begin{table*}
  \caption{\label{tab:bf}
    Partial and total branching fractions, in units of $10^{-4}$,
    measured with the semileptonic and hadronic
    tag analyses.
    The $q^2$ ranges are in $\gev^2$.
    The errors are statistical and systematic.
    The combined results are expressed as $\Bztopimlnu$ branching fractions.}
  \begin{tabular}{llccccr}
    \hline\hline
    & & $q^2<8$ & $8<q^2<16$ & $q^2>16$
      & $q^2<16$
      & \multicolumn{1}{c}{Total} \\
    \hline
    $\Bz$
    & Semileptonic
      & $0.50\pm0.16\pm0.05$ & $0.33\pm0.14\pm0.04$ & $0.29\pm0.15\pm0.04$
      & $0.83\pm0.22\pm0.08$ & $1.12\pm0.25\pm0.10$ \\
    & Hadronic
      & $0.09\pm0.10\pm0.02$ & $0.33\pm0.15\pm0.05$ & $0.65\pm0.20\pm0.13$ 
      & $0.42\pm0.18\pm0.05$ & $1.07\pm0.27\pm0.15$ \\
    & Average
      & $0.38\pm0.12\pm0.04$ & $0.33\pm0.10\pm0.03$ & $0.47\pm0.13\pm0.06$ 
      & $0.72\pm0.16\pm0.06$ & $1.19\pm0.20\pm0.10$ \\
    $\Bp$
    & Semileptonic
      & $0.18\pm0.08\pm0.02$ & $0.45\pm0.13\pm0.05$ & $0.10\pm0.12\pm0.04$
      & $0.63\pm0.16\pm0.06$ & $0.73\pm0.18\pm0.08$ \\
    & Hadronic
      & $0.16\pm0.11\pm0.03$ & $0.39\pm0.16\pm0.06$ & $0.26\pm0.12\pm0.06$ 
      & $0.56\pm0.19\pm0.08$ & $0.82\pm0.22\pm0.11$ \\
    & Average
      & $0.18\pm0.07\pm0.02$ & $0.43\pm0.10\pm0.04$ & $0.22\pm0.09\pm0.05$ 
      & $0.61\pm0.12\pm0.05$ & $0.82\pm0.15\pm0.09$ \\
    \multicolumn{2}{l}{Combined}
      & $0.36\pm0.09\pm0.03$ & $0.52\pm0.10\pm0.04$ & $0.46\pm0.10\pm0.06$ 
      & $0.87\pm0.13\pm0.06$ & $1.33\pm0.17\pm0.11$ \\
    \hline\hline
  \end{tabular}
\end{table*}
The largest sources of systematic error~\cite{epaps} are:
  the $\Btag$ reconstruction efficiency (discussed above),
  the shape of the background $\RR$ distribution
  (studied with control samples that fail the signal selection criteria),
  and the branching fractions of the $B$ semileptonic decays
  other than $B\to\pi\ell\nu$
  (varied within the current knowledge~\cite{pdg2004}).


In the second analysis, 
  we reconstruct the $\Btag$ meson in a set of purely 
  hadronic final states $B\to\DorDstar X$.
We reconstruct $\Dz$ mesons in
	$\Km\pip$, $\Km\pip\piz$, $\Km\pip\pip\pim$, and $\KS\pip\pim$ decays,
	and $\Dp$ mesons in $\Km\pip\pip$, $\Km\pip\pip\piz$, $\KS\pip$,
	$\KS\pip\piz$, and $\KS\pip\pip\pim$ decays.
The $\Dstar$ mesons are reconstructed in $\Dz\pip$, $\Dz\piz$,
	and $\Dz\gamma$ decays.
The hadronic system $X$ has a total charge $\pm1$ and is composed of
  $n_1\pipm + n_2\Kpm + n_3\piz + n_4\KS$ where
  $n_1+n_2<6$, $n_3<3$ and $n_4<3$.
The total reconstruction efficiency for a $\Bz$ ($\Bp$) meson is
  0.3\% (0.5\%).


We separate correctly-reconstructed $\Btag$ mesons from the background
  using two kinematic variables:
  the beam-energy substituted mass
  $\mES = \sqrt{s/4-|\pB|^2}$ and
  the energy difference
  $\DeltaE = E_B-\sqrt{s}/2$, where
  $\sqrt{s}$ is the c.m.\ energy of the $\epem$ system.
We select signal candidates in mode-dependent $\DeltaE$ windows
  around zero.
We apply a loose selection $5.2<\mES<5.3\gev$ 
  and fit the $\mES$ distribution at a later stage
  to extract the signal yield.


After reconstructing the $\Btag$, we look for the signature of
  a $\Btopilnu$ decay in the recoiling system.
The selection criteria for the pion and lepton candidates are
  similar to the first analysis, except
  a) the minimum $|\pell|$ for electrons is $0.5\gev$, and
  b) the $\piz$ mass window is 110--$160\mev$.
We require $\Eres<450\mev$
  for the $\Bz$ channel to reduce the $\Bz\to\rhom\ellp\nu$
  background, and no requirement is made for the $\Bp$ channel.

\begin{figure}
  \epsfig{file=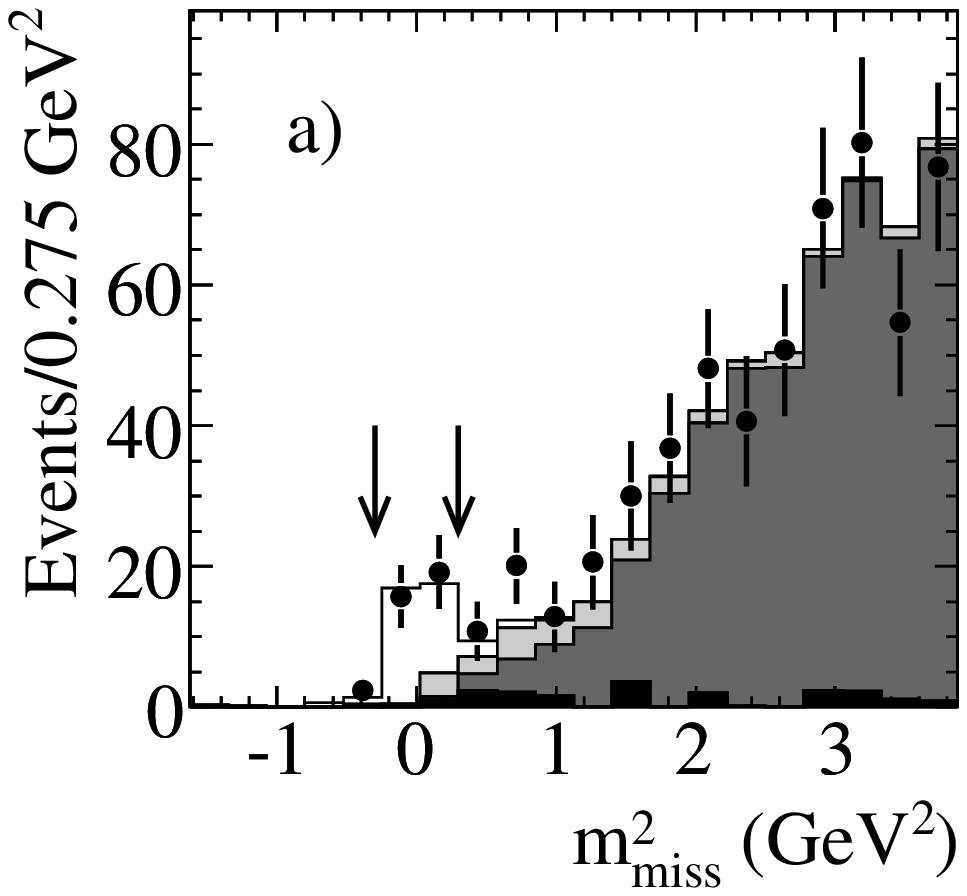,width=0.5\columnwidth}%
  \epsfig{file=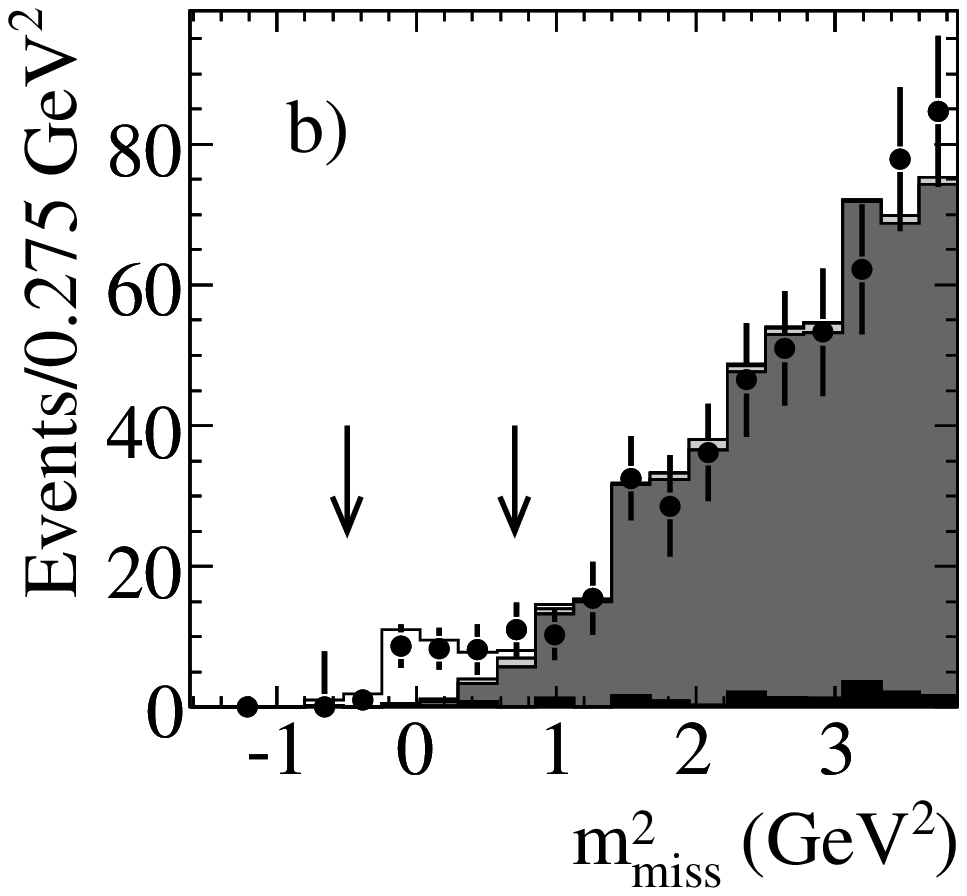,width=0.5\columnwidth}
\caption{\label{fig:mmiss}
	Distributions of $\mmiss$ of the
	a) $\Bz\to\pim\ellp\nu$ and b) $\Bp\to\piz\ellp\nu$ candidates.
	The points with error bars are the data.
	The histograms represent, from the lightest to the darkest,
	the MC simulation of the $B\to\pi\ell\nu$ signal, $b\to u\ell\nu$,
	$b\to c\ell\nu$, and other backgrounds.
	The arrows indicate the regions in which the signals are extracted.
}
\end{figure}

The full reconstruction of $\Btag$ allows us to determine the
  neutrino four-momentum precisely from the missing four-momentum
  $\pmiss=p_{\Upsilon(4S)}-p_{\Btag}-p_{\pi}-p_{\ell}$.
The missing mass squared $\mmiss$ peaks near zero for the signal
  and extends above zero for the background
  (see Fig.~\ref{fig:mmiss}).
We require $|\mmiss|<0.3\gev^2$ for the $\Bz$ channel and
  $-0.5<\mmiss<0.7\gev^2$ for the $\Bp$ channel, with the
  latter being broader and asymmetric due to the 
  resolution of the $\piz$ energy measurement.


Precise knowledge of $\pmiss$ allows us
  to calculate $q^2$ with small uncertainties.
We divide the signal candidates into the same three $q^2$ bins as before,
  and subtract the small bin-to-bin migration as background.
In each $q^2$ bin, we obtain the number of correctly-tagged events
  by an unbinned maximum likelihood fit
  to the $\mES$ distribution.
The PDF for the signal is determined from MC simulation
  as a Gaussian function joined to an exponential tail.
For the background, we use a threshold function of the form
  $x\sqrt{1-x^2}\exp(-\xi(1-x^2))$,
  where $x=2\mES/\sqrt{s}$ and the parameter $\xi$ is allowed
  to float in the fit.
Fig.~\ref{fig:mmiss} shows the $\mmiss$ distribution obtained by
  splitting the data samples in bins of $\mmiss$ and repeating
  the $\mES$ fit.

The signal side of the correctly-tagged events may not be
    a $\Btopilnu$ decay.
Contributions from this type of background are estimated with
    the MC simulation,
    as indicated by shaded histograms in Fig.~\ref{fig:mmiss},
	which are scaled to match the data in the sideband region
  	$1<\mmiss<4\gev^2$.
After background subtraction,
	we find signal yields of $31\pm7$ events and $26\pm7$ events
	for the $\Bz$ and $\Bp$ channels, respectively, 
	where the errors are statistical.


Instead of estimating the absolute signal efficiency,
  we normalize the signal yield to the number of inclusive
  $B$ semileptonic decays, $B\to X\ell\nu$, in the recoil
  of $\Btag$.
The reconstruction efficiencies of the $\Btag$ and of the lepton 
  cancel to first order in the ratio between the yields
  of the signal and normalization samples.
The inclusive branching fraction $\BR(B\to X\ell\nu)$
  is taken as $10.73\pm0.28\%$~\cite{pdg2004}.
The yield of the normalization sample is extracted by a fit
  to the $\mES$ distribution.
The component of the background that peaks in the $\mES$
  distribution is estimated from the MC simulation and 
  subtracted.
Efficiency differences between the signal and normalization
	samples are estimated with the MC
  simulation, and the corresponding corrections are applied to
  the result.


The measured branching fractions are summarized in Table~\ref{tab:bf}.
The largest source of systematic error
   is the limited statistics of the signal MC sample.
Other significant sources include
  the modeling of the signal PDF
  (studied with alternative fitting methods),
  photon-energy measurement,
  $\piz$ reconstruction,
  muon identification, and
  the branching fractions of non-signal $B\to X_u\ell\nu$ decays.


We take weighted averages of the measured partial branching fractions
    in each $q^2$ bin.
The results for the $\Bz$ and $\Bp$ channels
	are consistent with the isospin relation
	$\Gamma(\Bz\to\pim\ellp\nu) = 2\Gamma(\Bp\to\piz\ellp\nu)$
	and the lifetime ratio 
	$\tauBp/\tauBz = 1.081\pm0.015$~\cite{pdg2004},
	with $\chi^2 = 5.2$ for 3 degrees of freedom.
Assuming isospin symmetry, we combine the $\Bz$ and $\Bp$ channels
	and express the results as the $\Bz$ branching fraction
  in the last row of Table~\ref{tab:bf}.
The overall $\chi^2$ is 10.2 for 9 degrees of freedom.


We extract $\Vub$ from the partial branching fractions $\DBF$ using
  $\Vub = \sqrt{\DBF/(\tauBz\Delta\zeta)}$,
  where $\tauBz = (1.536\pm0.014)\ps$~\cite{pdg2004} is
  the \Bz\ lifetime and $\Delta\zeta = \Delta\Gamma/\Vub^2$
  is the normalized partial decay rate predicted by the
  form-factor calculations.
We use the light-cone sum rules calculation~\cite{lcsr:pi}
  for $q^2<16\gev^2$ and the
  lattice QCD calculations~\cite{lqcd:hpqcd,lqcd:fnal,lqcd:ape}
  for $q^2>16\gev^2$.
The results are shown in Table~\ref{tab:vub}.


In conclusion, we have measured the $\Btopilnu$ branching fraction
  as a function of $q^2$ using tagged $B$ meson samples,
  and have extracted $\Vub$.
The measured total branching fraction,
  $\BR(\Bz\to\pim\ellp\nu) = (1.33\pmstat{0.17}\pmsyst{0.11})\times10^{-4}$,
  has the smallest systematic uncertainty among the existing
  measurements~\cite{cleo_vub,babar_vub,belle_vub} thanks to the
  superior signal purity, and the overall precision is comparable to the
  best.
Using theoretical calculations of the form factor,
  we obtain values of $\Vub$ ranging
  between $3.2\times10^{-3}$ and $4.5\times10^{-3}$.
As an example,
  the recently published unquenched lattice QCD calculation~\cite{lqcd:hpqcd}
  gives $\Vub = (4.5\pmstat{0.5}\pmsyst{0.3}\pmff{0.7}{0.5})\times10^{-3}$.
Improvement will be possible with additional data
  combined with more precise form-factor calculations.
  
\begin{table}
  \caption{\label{tab:vub}
    Values of $\Vub$ derived using the form factor calculations.
    The first two errors on $\Vub$ come from the statistical and systematic
    uncertainties of the partial branching fractions.
    The third errors correspond to the uncertainties on $\Delta\zeta$
    due to the form-factor calculations, and are taken from 
    Refs.~\onlinecite{lcsr:pi,lqcd:hpqcd,lqcd:fnal,lqcd:ape}.
  }
  \begin{tabular}{lccc}
  \hline\hline
  & $q^2$ (GeV$^2$) & $\Delta\zeta$ (ps$^{-1}$) & $\Vub$ ($10^{-3}$) \\
  \hline
  Ball-Zwicky~\cite{lcsr:pi} & $<16$ & $5.44\pm1.43$ & $3.2\pm0.2\pm0.1\err{0.5}{0.4}$ \\
  Gulez \etal~\cite{lqcd:hpqcd} & $>16$ & $1.46\pm0.35$ & $4.5\pm0.5\pm0.3\err{0.7}{0.5}$ \\
  \makebox[0pt][l]{Okamoto \etal~\cite{lqcd:fnal}} & $>16$ & $1.83\pm0.50$ & $4.0\pm0.5\pm0.3\err{0.7}{0.5}$ \\
  Abada \etal~\cite{lqcd:ape} & $>16$ & $1.80\pm0.86$ & $4.1\pm0.5\pm0.3\err{1.6}{0.7}$ \\
  \hline\hline
  \end{tabular}
\end{table}

We are grateful for the excellent luminosity and machine conditions
provided by our \pep2\ colleagues, 
and for the substantial dedicated effort from
the computing organizations that support \babar.
The collaborating institutions wish to thank 
SLAC for its support and kind hospitality. 
This work is supported by
DOE
and NSF (USA),
NSERC (Canada),
IHEP (China),
CEA and
CNRS-IN2P3
(France),
BMBF and DFG
(Germany),
INFN (Italy),
FOM (The Netherlands),
NFR (Norway),
MIST (Russia),
MEC (Spain), and
PPARC (United Kingdom). 
Individuals have received support from the
Marie Curie EIF (European Union) and
the A.~P.~Sloan Foundation.

\bibliography{references}

\begin{thebibliography}{12}
\bibitem[{cha()}]{charge}
\bibinfo{note}{Charge conjugate states are implied throughout this Letter.}
\end{thebibliography}


\setcounter{table}{0}

\begin{table*}[h]
\flushleft
\textbf{\large Electronic Physics Auxiliary Publication Service (EPAPS)}\\
\smallskip
\normalsize
This is an EPAPS attachment to
B.~Aubert \textit{et al.} (\babar\ Collaboration),
Phys.\ Rev.\ Lett.\ \textbf{97}, 211801 (2006)
[arXiv:hep-ex/0607089].
For more information on EPAPS, see
\url{http://www.aip.org/pubservs/epaps.html}.
\end{table*}

\begin{table*}[h]
\centering
\caption{\label{tab:syst}
  Fractional systematic errors (in \%) of the measured partial 
  branching fractions.
  The $q^2$ bins are defined as
  1: $q^2<8\gev^2$, 2: $8<q^2<16\gev^2$, and 3: $q^2>16\gev^2$.
  The $+$ and $\times$ symbols indicate if the error is additive (+) or 
  multiplicative ($\times$).}
\begin{tabular}{lc|rrr|rrr|rrr|rrr}
\hline\hline
\multicolumn{2}{r|}{$\Btag$} 
  & \multicolumn{3}{c|}{$\Bz$ semilep.} & \multicolumn{3}{c|}{$\Bp$ semilep.}
  & \multicolumn{3}{c|}{$\Bz$ hadronic} & \multicolumn{3}{c}{$\Bp$ hadronic} \\
\multicolumn{2}{r|}{bin}
  & 1 & 2 & 3  & 1 & 2 & 3  & 1 & 2 & 3  & 1 & 2 & 3 \\
\hline
       $\Btopilnu$ form factor & $\times$ & $1.0$ & $0.5$ & $1.1$ & $0.9$ & $0.5$ & $ 4.5$ & $ 0.3$ & $ 0.2$ & $ 0.1$ & $ 0.3$ & $ 0.2$ & $ 2.2$ \\
        $\BtoXclnu$ background & $+$ & $ 1.9$ & $ 2.9$ & $ 3.8$ & $ 2.0$ & $ 3.5$ & $ 7.7$ & $ 0.2$ & $ 0.2$ & $ 0.2$ & $ 2.6$ & $ 2.6$ & $ 2.6$ \\
        $\BtoXulnu$ background & $+$ & $ 0.8$ & $ 1.7$ & $ 6.9$ & $ 1.2$ & $ 1.7$ & $12.1$ & $ 4.2$ & $ 4.2$ & $ 4.2$ & $ 1.7$ & $ 1.7$ & $ 1.7$ \\
               $\BR(\BtoXlnu)$ & $\times$ & \multicolumn{6}{c|}{not applicable} & $ 2.6$ & $ 2.6$ & $ 2.6$ & $ 2.6$ & $ 2.6$ & $ 2.6$ \\
 $\BR(\Upsilon(4S)\to\Bz\Bzb)$ & $\times$ & $ 1.6$ & $ 1.6$ & $ 1.6$ & $ 1.6$ & $ 1.6$ & $ 1.6$ & \multicolumn{6}{c}{not applicable} \\
         Final-state radiation & $\times$ & $ 1.2$ & $ 1.2$ & $ 1.2$ & $ 1.2$ & $ 1.2$ & $ 1.2$ & $1.2$ & $1.2$ & $1.2$ & $1.2$ & $1.2$ & $1.2$ \\
            $\Btag$ efficiency & $\times$ & $ 7.3$ & $ 7.3$ & $ 7.3$ & $ 4.3$ & $ 2.5$ & $12.9$ & $ 0.7$ & $ 0.7$ & $ 0.7$ & $ 1.4$ & $ 1.4$ & $ 1.4$ \\
              $q^2$ resolution & $\times$ & $ 1.6$ & $ 1.3$ & $ 1.2$ & $ 1.2$ & $ 4.5$ & $18.0$ & \multicolumn{6}{c}{negligible} \\
                    Fit method & $+$ & $ 1.4$ & $ 2.1$ & $ 5.7$ & $ 4.8$ & $ 6.2$ & $32.8$ & $ 5.7$ & $ 5.7$ & $ 5.7$ & $ 2.7$ & $ 2.7$ & $ 2.7$ \\
         Lepton identification & $\times$ & $ 1.6$ & $ 1.9$ & $ 1.9$ & $ 2.5$ & $ 2.5$ & $ 2.5$ & $ 2.5$ & $ 2.5$ & $ 2.5$ & $ 2.5$ & $ 2.5$ & $ 2.5$ \\
  Charged track reconstruction & $\times$ & $ 1.6$ & $ 1.6$ & $ 1.6$ & $ 0.8$ & $ 0.8$ & $ 0.8$ & $ 1.1$ & $ 1.1$ & $ 1.1$ & $ 1.4$ & $ 1.4$ & $ 1.4$ \\
 Neutral energy reconstruction & $\times$ & \multicolumn{3}{c|}{negligible} & $ 3.2$ & $ 3.0$ & $ 6.3$ & $ 1.2$ & $ 1.2$ & $ 1.2$ & $ 3.7$ & $ 3.7$ & $ 3.7$ \\
         Number of \BB\ events & $\times$ & $ 1.1$ & $ 1.1$ & $ 1.1$ & $ 1.1$ & $ 1.1$ & $ 1.1$ & \multicolumn{6}{c}{not applicable} \\
                 MC statistics & $\times$ & $ 5.2$ & $ 5.1$ & $ 4.6$ & $ 4.7$ & $ 4.2$ & $ 6.8$ & $18.3$ & $11.8$ & $17.6$ & $19.8$ & $14.7$ & $23.0$ \\
\hline
                         Total & & $10.0$ & $10.4$ & $13.6$ & $ 9.7$ & $10.9$ & $43.5$ & $20.1$ & $14.4$ & $19.4$ & $21.0$ & $16.3$ & $24.1$ \\
\hline\hline
\end{tabular}
\end{table*}

\begin{table*}[h]
\centering
\caption{\label{tab:matrix}
  Values and errors (in unit of $10^{-4}$) of the combined partial
  branching fractions.
  The errors are separated into statistical, multiplicative systematic,
  and non-multiplicative systematic components, and the covariance 
  matrices for the systematic components are given.
  The $q^2$ bins are defined as
  1: $q^2<8\gev^2$, 2: $8<q^2<16\gev^2$, and 3: $q^2>16\gev^2$.}
\begin{tabular}{lc|ccc}
\hline\hline
& bin & 1 & 2 & 3  \\
\hline
Value              & & 0.355 & 0.518 & 0.457 \\
\hline
Statistical error  & & 0.086 & 0.097 & 0.104 \\
\hline
Multiplicative systematic error & & 0.028 & 0.038 & 0.052 \\
\hline
Covariance   & 1 & 1.000 & 0.593 & 0.471 \\
             & 2 & 0.593 & 1.000 & 0.504 \\
             & 3 & 0.471 & 0.504 & 1.000 \\
\hline
Non-multiplicative systematic error & & 0.008 & 0.017 & 0.021 \\
\hline
Covariance   & 1 & 1.000 & 0.986 & 0.791 \\
             & 2 & 0.986 & 1.000 & 0.881 \\
             & 3 & 0.791 & 0.881 & 1.000 \\
\hline\hline
\end{tabular}
\end{table*}

\end{document}